\documentclass{cjp}

\volyear{916}{2006}

\received{October 1, 2006}

\accepted{October 30, 2006}

\begin{document}

\title{Correlations in Many Electron Systems: Theory and Applications }

\author[M. Tomaselli]{M. Tomaselli}

\address{TUD, Technical University Darmstadt, D64289 Darmstadt, Germany and
          GSI-Gesellschaft f{\"u}r Schwerionenforschung, D64291 Darmstadt, Germany.}    

\author[T. K{\"u}hl]{T. K{\"u}hl}

\address{ GSI-Gesellschaft f{\"u}r Schwerionenforschung, D64291 Darmstadt, Germany.}    
 
\author[D. Ursescu]{D. Ursescu}

\address{ GSI-Gesellschaft f{\"u}r Schwerionenforschung, D64291 Darmstadt, Germany.}

\author[S. Fritzsche]{S. Fritzsche}

\address{Institute of Physics, Kassel University, D34132 Kassel, Germany}  

\shortauthor{Tomaselli, K{\"u}hl, Ursescu, and Fritzsche}

\maketitle

\begin{abstract}
In this contribution we present calculations performed for interacting electron systems 
within a non-perturbative formulation of the cluster theory.
Extrapolation of the model to describe the time dependence 
of the interacting systems is feasible and planed. 
The theory is based on
the unitary operator $e^{iS}$ ({\it S} is the correlation operator) formalism which, 
in this paper, is treated non perturbatively within many-particle correlations.
The application of the derived equations to few-body systems is realized in
terms of Generalized Linearization Approximations (GLA) and
via the Cluster Factorization Theory (CFT).  
To check the reliability of the model we 
present two different applications.
In the first we evaluate the transitions energies in Helium-,
Lithium-, Beryllium-, and Boron-like Oxygen.
The calculation aims to a precise determination of the satellite transitions
which play an important role in plasma diagnostics.
In a second we investigate a non-perturbative method to evaluate the charge radii of the
Helium and Lithium isotopes by using the Isotopic Shift theory.  
We have found that our model leads naturally to components of $e^--e^+$ pair
in the two-electron wave functions of the Helium
isotopes and three-electron wave functions of the Lithium isotopes.
The possible connection of these terms to the QED leading diagrams is
postulated.


\end{abstract}

\begin{resume}

French version of abstract (supplied by CJP)

   \traduit

\end{resume}

\section{Introduction}
Deriving a non-perturbative and microscopic theory capable
to describe the basic observable that characterize the dynamics of interacting electrons
is a fundamental problem in the physics of atoms and ions.
In general, one faces with two fundamental tasks, namely,    
the consideration of the correlation effects and
the introduction of a cut-off parameter
which, in order to obtain
 realistic and solvable systems, reduces the dimensions of the model Equation of Motion (EoM).
The introduction of correlation effects in
many body systems via the $e^{iS}$ Unitary-Model Operator (UMO) goes back
to the early work of Villars~\cite{vil61}.
The idea is to introduce a wave operator S which maps zero-order reference
wave functions (usually Hartree-Fock wave functions) to exact many body wave functions.    
Extended applications of the method in nuclear physics were shortly after
performed by Shakin~\cite{sha01}.
The $e^{iS}$ method came to quantum chemistry with the coupled cluster method
proposed by Coester~\cite{coe58}, and K{\"u}mmel~\cite{kum60}.
The coupled cluster Hamiltonian has been recently applied to the calculations
of the electron affinities of alkali atoms~\cite{eli05}.
Studies of correlation effects in atomic systems based on
the coupled cluster theory have been performed by Das et al.~\cite{das05}.
Recently~\cite{tom01,tom02} the $e^{iS}$ method was applied within 
nonperturbative approximations (Dynamic Correlation Model (DCM) 
and Boson Dynamic Correlation Model (BDCM)) to open shell nuclei.
Applications of the method to open-shell electron systems were firstly applied   
to calculate the Hyperfine Splitting (HFS) constants of Lithium-like 
bismuth and uranium~\cite{tom04,tom06}. 
The resulting non-perturbative and relativistic electron Dynamic Correlation Model (eDCM)
 was applied to calculate the effect produced by the electron and 
nucleon correlations into the isotopic 
shift theory IS. Calculations for lithium atoms were presented in~\cite{tom05}.   
Additionally the method finds application in the evaluation of 
dielectronic satellite-spectra of Lithium-like ions~\cite{iyu97,ros02,pik96}.
 These are a useful tool for diagnostic of laser produced plasma.
The ratio of various components of the satellite lines  have been shown to be sensitive 
to density and temperature. 

We start by describing free electron systems with
a relativistic shell model in which the wave functions are 
solution of the Dirac's equation.
The model vacuum consists in paired electrons to fill major shells.
The electrons in excess are considered as valence particles.
The interaction between the electrons is responsible for exciting the valence
electrons and for causing correlation effects in the closed shells.
In additions to this polarization mechanism we have also the polarization of
the continuum states. This polarization effects
 named Boiling of the Vacuum (BoV), have been already introduced
in \cite{tom04}.  
 As in Ref.~[7] we start by defining the basic operators of the model and 
by determining the relative EoM. 
The complex excitations modes are classified in terms of electron
Configuration Mixing Wave Functions (eCMWFs).
The eCMWFs form  an orthogonal base of coupled clusters in which the Pauli 
principle between the different clusters is taken fully in consideration.
Extrapolation of the non-perturbative cluster model to describe the time 
dependent electron-laser interaction is feasible and planed. 

In this contribution we present two applications of the non perturbative eDCM.
The first involves the evaluation of the relativistic transition energies and
wave functions for the Oxygen ions ranging from the Helium-like
 to the Boron-like. 
In the second application we study the dynamics of few-electron systems interacting
with the excitation of the positron-continuum.  
The effect of this excitations is important in the determination of a non perturbative
descriptions of the Mass Shift (MS) and Field-Shift (FS) which characterize 
the Isotopic Shift (IS) theory.

\section{Theory}
We start with a set of exact eigenstates $\{|\nu\rangle \}$ of the Dirac's Hamiltonian:
\begin{equation}\label{equ.2.0}
\displaystyle{
h_i= c\vec{\alpha_i}\vec{p_i}+(\beta-1)+v_{\mathrm{nucl}}(r_i)}
\end{equation}
 which satisfies the dynamical equation 
\begin{equation} \label{eq11}
 H|\nu \rangle = E_{\nu}|\nu\rangle\ .
\label{equ.2.1} 
\end{equation}   
In dealing with many electron systems one has to add the correlation effects 
caused by the two-body interactions:
$V(ij)_{\mathrm{Coul}}$ and $V(ij)_{\mathrm{Breit}}$ to the Hamiltonian of Eq.~(\ref{equ.2.0}).
Shell model calculation can be then performed to calculate transition energies
between the different levels. Shell model calculations
represent however an approximation in that one usually treats the effects of only
few shells. The neglected shells serve to re-normalize the interaction
 in the shells considered.
The re-normalization of the Hamiltonian is generally introduced via
correlation operators. 
In UMO the effective Hamiltonian is calculated by introducing
the correlations via the unitary $e^{iS}$ operator.
By using only two body correlation we can derive:
\begin{equation}\label{eq.1}
\begin{array}{l}
\displaystyle{
H_{eff}=e^{-iS_2}He^{iS_2}=\sum_{\alpha\beta}\langle\alpha|t|\beta\rangle 
a^{\dagger}_{\alpha}a_{\beta}+
\sum_{\alpha\beta\gamma\delta}\langle\Psi_{\alpha\beta}|v_{12}|\Psi_{\gamma\delta}\rangle a^{\dagger}_{\alpha}a^{\dagger}_{\beta} a_{\delta}a_{\gamma}}\\
\displaystyle{
=\sum_{\alpha\beta}\langle\alpha|t|\beta\rangle a^{\dagger}_{\alpha}a_{\beta}+
\sum_{\alpha\beta\gamma\delta}\langle\Psi_{\alpha\beta}|v|\Psi_{\gamma\delta}\rangle a^{\dagger}_{\alpha}a^{\dagger}_{\beta} a_{\delta}a_{\gamma}}
\end{array}
\end{equation}
where $v_{12}$ is the two body interaction 
 and the $\Psi_{\alpha\beta}$ is the two particle correlated wave function:
\begin{equation}\label{eq.1a}
\Psi_{\alpha\beta}=e^{iS_2}\Phi_{\alpha\beta}
\end{equation}

However in dealing with complex atoms the ($S_i,~i=3\cdots n$)
correlations should also be considered.
The evaluation of these diagrams is, due to the 
exponentially increasing number of terms, difficult in a perturbation theory.

We note that one way to overcome this problem is to work with 
$e^{i(S_1+S_2+S_3+\cdots+S_i)}$ operator on the Slater's determinant of the 
different states by keeping the {\it n}-body Hamiltonian uncorrelated.

After having performed the diagonalization of eigenvalue matrix obtained
from the matrix elements of
the n-body uncorrelated Hamilton's operator, we
can calculate the form of the effective Hamiltonian which, by now, includes 
correlation operators of complex order.

The amplitudes of the correlated determinant are the calculated in
the EoM method which is illustrated in the following.

If $|0\rangle$ denotes some physical vacuum  
and $O^{\dagger}_{\nu}$ denotes the operator that creates 
the many-body eigenstate $|\nu\rangle$ such that   
$O^{\dagger}_{\nu}|0\rangle = |\nu\rangle$,   
$O_{\nu}|0\rangle= 0$, and $H|0\rangle = E_0|0\rangle$, then we have  
a set of EoM of the form   
\begin{equation}
i\hbar \frac{\partial O^{\dagger}_{\nu}}{\partial t} |0\rangle
= [H, O^{\dagger}_{\nu}]|0\rangle = (E_{\nu}-E_0)|\nu\rangle \equiv 
\omega_{\nu}O^{\dagger}_{\nu}|0\rangle \ . 
\label{equ.2.2} 
\end{equation} 
In terms of the operators, the EoM can be written as   
\begin{equation} 
  [H, O^{\dagger}_{\nu}] = \omega_{\nu}O^{\dagger}_{\nu} \ . 
\label{equ.2.3} 
\end{equation} 
In Eq.~(\ref{equ.2.3}) the Hamiltonian has the general second quantization form   
\begin{equation}\label{f2}
\begin{array}{l}
H=\sum_{\alpha}\epsilon_{\alpha} c^{\dagger}_{\alpha}c_{\alpha}+\frac{1}{2}
\sum_{\alpha\beta\delta\gamma} \langle \alpha\ \beta|v(r)|\delta\ \gamma\rangle 
c^{\dagger}_{\alpha} c^{\dagger}_{\beta}c_{\delta}c_{\gamma}\\
=T+V_{\mathrm{int}}
\end{array}
\end{equation} 
where T is the kinetic energy operator and $V_{\mathrm{int}}$
the interactions ($V_{\mathrm{Coul}}$+$V_{\mathrm{Breit}}$), and the $c^{\dagger}, c$ 
the general fermion operators.  
When they act on valence subspace, the $c^{\dagger}$ and $c$ creates and 
annihilates a valence electron, respectively. On the other hand, 
when they act on core subspace, the $c^{\dagger}$ and $c$ respectively
annihilates and creates a hole state.     
Hence, the summation of the Greek subscripts leads to    
particle-particle, particle-hole, as well as   
hole-hole interactions. 

It is useful to determine the form of the central potential
 before the diagonalization of the model space is performed. This is because  
the matrix elements of the EoM can often be more easily calculated in a 
pre-diagonalization basis.   
  
If $\{|b\rangle\} ( = |1\rangle, |2\rangle, \ldots, |r\rangle, \ldots) $ 
is a complete set of basis vectors, 
then  
\begin{equation} 
|\nu \rangle 
= \sum_{b} |b\rangle\langle b|\nu\rangle \equiv \sum_{b} x_{b\nu }\ |b\rangle \ , 
\label{equ.2.4} 
\end{equation} 
or  
\begin{equation}
O^{\dagger}_{\nu} = \sum_{b} x_{b \nu}\ O^{\dagger}_{b} \ . 
\label{equ.2.5}
\end{equation} 
Using this last relation in Eq.~(\ref{equ.2.3}), we obtain  
\begin{equation}
 [H,O^{\dagger}_{r}] = \sum_{b} \Omega_{br} O^{\dagger}_{b} 
\label{equ.2.6}
\end{equation}
where 
$\Omega_{br}\equiv\sum_{\nu} x_{b\nu}\ \omega_{\nu}\ x^{-1}_{\nu r} \ .$      
Eq.~(\ref{equ.2.6}) is the general form of EoM for the operator $O_{\nu}$.  

The coefficients $\Omega_{br}$ are simply the matrix elements of 
the Hamiltonian. To see this, we take  
the matrix element of both sides of Eq.~(\ref{equ.2.6}) between the    
states $\langle s|$ and $|0\rangle$. Upon using the orthogonality between
the basis vectors ({\em i.e.}    
$\langle s|
O^{\dagger}_r|0\rangle = \delta_{rs}$), 
one obtains  
\begin{equation} 
\Omega_{sr} = \langle s| H | r \rangle - E_0\delta_{sr} \ .      
\label{equ.2.6b} 
\end{equation} 

If the model space consists of a finite number, $N$, 
of basis vectors, then going 
from Eq.~(\ref{equ.2.6}) back to Eq.~(\ref{equ.2.3}) is  
equivalent to associate the systems of coupled equations given in Eq.~(\ref{equ.2.6b})
 to the eigenvalues matrix equation given below:
\begin{equation}
( {\bf O}\ - E\ {\bf 1})\  {\bf x} = 0  
\label{equ.2.6c}
\end{equation} 
where ${\bf O}$ represents the $(N\times N)$ matrix $\Omega$, \ 
${\bf 1}$ the $N$-dimensional unit matrix, and $\bf x$
are the projections of the model space into the basic vectors.  

Equations (\ref{equ.2.6}) and (\ref{equ.2.6b}) indicate that 
the complexity of solving Eq.~(\ref{equ.2.6c}) depends on the 
complexity of the model space, $\{|b\rangle\}$,  and the 
Hamiltonian, $H$. The following comparative review of the construction of 
model spaces in different structure theories should give a 
glimpse on the scope of the problem.
   
Let $O^{\dagger}_m$ be the operator  
that creates $n$ valence electrons outside the closed shells 
 state $|\Phi_0\rangle$ :   
\begin{equation}\label{f1}
|m\rangle \equiv O^{\dagger}_m(\alpha_m;j_1j_2\cdots j_n)= \prod_{i=1}^{n}\ a^{\dagger}_{j_i}|\Phi_0\rangle    
= |\alpha_m; j_1j_2\cdots j_n\rangle \ .  
\end{equation}
In the simplest case where there is no closed shell excitation, 
the $O_m^{\dagger}$ satisfies the EoM, Eq.~(\ref{equ.2.6})       
\begin{equation}\label{f3}
[H,O^{\dagger}_m(\alpha_m;j_1j_2\cdots j_n)] =\sum_{m'}\Omega_{m m'}
O^{\dagger}_{m'}(\alpha_{m'};j'_1j'_2\cdots j'_n) \ ,  
\end{equation}
with $\alpha_m$ and $\alpha_{m'}$ denoting the quantum numbers of the 
states $|m\rangle$ and   
$|m'\rangle$, respectively.     

The inert-core approximation would be good 
only if the valence-core interaction is very small.   
Hence, the applicability of the inert-core approximation is  
very limited as the    
interaction between valence and core electrons will generally excite the 
shell-model ground state of the core and create, in the process, the  
particle-hole ($ph$) pairs. Inclusion of the excitation mode 
due to $1p1h$ in the model space is known as   
the Tamm-Dancoff approximation (TDA)~\cite{bro64}. If one defines   
\begin{equation}
|m\rangle_{TDA} = A^{\dagger}_m\ |0\rangle_{TDA} \ ,  
\label{equ.2.8}  
\end{equation} 
then Eq.~(\ref{equ.2.5}) takes the form  
\begin{equation} 
A^{\dagger}_m = \sum_{m'}\left[ 
\sum_{j_1j_2}\chi_{j_1j_2}^{(m')}a^{\dagger}_{j_1}b^{\dagger}_{j_2}
\otimes  \chi^{(m')}_o O^{\dagger}_{m'}\right]_m  
\label{equ.2.7}
\end{equation}
The $b^{\dagger}_{j_2}$ creates a hole $j^{-1}_{2}$ in $|0\rangle_{TDA}$ 
by destroying a core electron of $j_2$ while 
$a^{\dagger}_{j_1}$ creates a valence electron of $j_1$. The $A^{\dagger}_m$
creates therefore a state of $n+1$ particles and $1$ hole (or   
$p^{n+1}h^{1}$).  
The $\chi$'s are the configuration mixing coefficients 
and $|0\rangle_{TDA}$ denotes the physical vacuum of the TDA. In the 
literature one often chooses   
$|0\rangle_{TDA}$ = $|HF\rangle$, with $|HF\rangle$ being
the Hartree-Fock ground state of the ion. In this latter case, 
$O^{\dagger}_{m'} = 1 $ in Eq.~(\ref{equ.2.7}).  

It is also possible to use a physical vacuum that already contains $ph$ pairs.
In the literature, the
method of random phase
approximation (EPA)~\cite{bro64} has been introduced to study the full effects due   
to the pre-existence of $1p1h$ component in the physical vacuum. 
Hence, in RPA    
\begin{equation}\label{f5} 
A^{\dagger}_m = 
\sum_{m'} \left[\ \sum_{j_1j_2} (\chi_{j_1j_2}^{(m')}a^{\dagger}_{j_1}b^{\dagger}_{j_2}
+\chi_{j_2j_1}^{(m')}b_{j_2}a_{j_1}) \otimes \chi_o^{(m')}O^{\dagger}_{m'}\right]_m   
\end{equation}
and   
\begin{equation}
|m\rangle_{RPA} = A^{\dagger}_m|0\rangle_{RPA} \ ,  
\label{equ.2.9}
\end{equation} 
one can see that the term $b_{j_2}a_{j_1}$  
gives a null result if the physical vacuum $|0_{RPA}\rangle$ 
does not contain  
pre-existing $ph$ pairs. 
(In the literature, the coefficients
$\chi_{j_1j_2}$ and $\chi_{J_2j_1}$ are denoted by $x^m_{j_1j_2}$
and $-y^m_{j_2j_1}$.) 
If the RPA is  
applied to closed-shell, then again $O^{\dagger}_{m'} = 1 $ in 
Eq.~(\ref{f5}). 
 
The introduction of the excitations of the vacuum in the above mentioned approximation
is however complicated by the fact that the TDA and RPA vacua are different
then the vacuum of the single particle operators. In addition simple calculations can be
performed only by prediagonalizing the many body Hamiltonian in the TDA and RPA subspaces.
The coupling to the additional valence particles can afterwards
be accomplished by considering only few collective states and 
by neglecting the full treatment of the Pauling principle.
In the following we show that these complications can be overcome by
extending the EoM method to the field of non-linear equations.  
\subsection{Polarization of the closed shells versus continuum vacuum excitations} 
In the eDCM, the model space is expanded to include multiple $ph$ 
excitations.  
This dynamic mechanism includes either the excitations of closed electron
shells or of positron-continuum states.
More specifically~\cite{tom02},  
the eDCM states are classified according to the number
of the valence electrons and of the electron particle-hole 
pair arising either from closed shells or from the positron-continuum.
A state of ${\it N}$ paired valence electrons and ${\it N'}$ particle-hole closed shells electrons or $e^--e^+$ positron-continuum states is defined by
\begin{equation}
|\Phi^{(N,N')}\rangle = A^{\dagger}_{(N,N') J}|0\rangle   
\label{equ.2.11}
\end{equation}
with 
\begin{equation}
\label{equ-ii-3} 
\begin{array}{c}
\displaystyle{ \: \: A^{\dagger}_{(N,N')J}  \: = \:  }
\displaystyle{ \left [ \sum_{\alpha_N(J_1 J_2 \ldots J_N)} X_{\alpha_N (J_1J_2\ldots J_N); J}
A^{\dagger}_N (\alpha_N (J_1 J_2 \ldots J_N); J) \right. } \\
\displaystyle{+  \sum_{\alpha_{N+1'}(J_1 J_2 \ldots J_{N+1'})}
X_{\alpha_{N+1'} (J_1 J_2 \ldots J_{N+1'}); J}
A^{\dagger}_{N+1'} (\alpha_{N+1'}( J_1 J_2 \ldots J_{N+1'}); J)\: } \\
\displaystyle{\left.+ \ldots + \sum_{\alpha_{N+N'}(J_1\ldots J_{N+N'})}
X_{\alpha_{N+N'} (J_1\ldots J_{N+N'}); J}
A^{\dagger}_{N+N'} (\alpha_{N+N'}(J_1 \ldots J_{N+N'}); J) \right]
},
\end{array}
\end{equation}
where ${\it J}$ denotes the total spin and the $\alpha'$s the other quantum
numbers. The  
unprimed indices $1, \ldots, n$ label the valence particle-particle 
pairs ( the valence bosons) and the primed indices $1', \ldots, n'$ label 
the particle-hole pairs (the core electrons).  
The $J_i$'s denote the coupling of the
pairs and the coupling of the different $J_i$ is for simplicity omitted.
The $X$'s are projections of the model states to the basic vectors of  
Eq.~(\ref{equ.2.11}). 

Within this definition the model space included either the excitation of the closed
shells or the dynamics of continuum excitation which is taken into account 
through coupling the valence electron states to $e^--e^+$ states.   
The electron states defined in Eq.~(\ref{equ.2.11}) are  
 classified in terms of configuration
mixing wave functions (eCMWFs) of increasing degrees of complexity (number of
particle-hole or of $e^--e^+$ pairs), see Ref.~\cite{tom01}. 

Since the different subspaces should be rotational invariant we introduce
the coupling of the particles and particle-holes in such a way that 
the first pair 
is coupled to angular momentum ${\it J_1}$, the second to ${\it J_2}$, 
the two pairs are then
coupled to ${\it J_3}$ and so on until all the pairs are coupled to the 
total angular momentum {\it J}, {\em e.g.},    
\begin{equation} 
\label{eq.f11}
\begin{array}{l}  
A^{\dagger}_{N}(\alpha_{N}(J_1J_2\cdots J_N);J)=\\ 
\displaystyle{\left [
 \left(\ [ \{\ (a^{\dagger}_1a^{\dagger}_2)^{J_1}(a^{\dagger}_3a^{\dagger}_4)^{J_2}\ \}^{\lambda_1}
(a^{\dagger}_5a^{\dagger}_6)^{J_3}\ ]^{\lambda_2}    
 \cdots \right)^{\lambda_{N-1}}(a^{\dagger}_{2n-1}a^{\dagger}_{2n})^{J_n} \right ]^J }   
\end{array}
\end{equation}
and 
\begin{equation}
\label{eq.11a}
\begin{array}{l}
A^{\dagger}_{N+1'}(\alpha_{N+1'}(J_1J_2\cdots J_{N+1'})J)=\\
\displaystyle{ \{ 
 \left(\ [ \{\ (a^{\dagger}_1a^{\dagger}_2)^{J_1}(a^{\dagger}_3a^{\dagger}_4)^{J_2}\ \}^{\lambda_1}
(a^{\dagger}_5a^{\dagger}_6)^{J_3}\ ]^{\lambda_2}
\cdots(a^{\dagger}_{2n-1}a^{\dagger}_{2n})^{J_n}\right)^{\lambda_N} } \\ 
\displaystyle{ (a^{\dagger}_{2n+1'}b^{\dagger}_{2n+2'})^{J_{n+1'}}  
 \} ^J } .  
\end{array}
\end{equation}
Introduction of Eq.~(\ref{equ-ii-3}) into Eq.~(\ref{equ.2.6}) gives the 
following equations of motion in the eDCM:  
\begin{equation}
\label{eq.9}
\begin{array}{l}
 \displaystyle{[H,A^{\dagger}_N(\alpha_N(J_1J_2\cdots J_N)J)]|0\rangle}\\
\displaystyle{ 
=\sum_{\beta_{N}} \Omega_{p^N\ p^N}\ A^{\dagger}_N(\beta_N(J_1J_2\cdots J_N)J)|0\rangle   }   \\
\displaystyle{+\sum_{\beta_{N+1'}} \Omega_{p^N\ p^{N+1}h^1}\ A^{\dagger}_{N+1'}(\beta_{N+1'}(J_1J_2\cdots J_{N+1'})J)|0\rangle } \\ 
\displaystyle{ + \cdots} 
\end{array}
\end{equation}
\begin{equation}
\label{eq.9a}
\begin{array}{l}
 \displaystyle{[H,A^{\dagger}_{N+1'}(\alpha_{N+1'}(J_1J_2\cdots J_{N+1'})J)]|0\rangle}\\
\displaystyle{
 =\sum_{\beta_{N}} \Omega_{p^{N+1}h^1\ p^N}\ A^{\dagger}_N(\beta_N(J_1J_2\cdots J_N)J)|0\rangle}\\
\displaystyle{+\sum_{\beta_{N+1'}} 
\Omega_{p^{N+1}h^1\ p^{N+1}h^1}\ A^{\dagger}_{N+1'}(\beta_{N+1'}(J_1J_2\cdots J_{N+1'})J)|0\rangle } \\ 
\displaystyle{+\sum_{\beta_{N+2'}} 
\Omega_{p^{N+1}h^1\ p^{N+2}h^2}\ A^{\dagger}_{N+2'}(\beta_{N+2'}(J_1J_2\cdots J_{N+2'})J)|0\rangle } \\ 
\displaystyle{ + \cdots} 
\end{array} 
\end{equation} 
where $|0\rangle$ is the shell-model state. 
Furthermore, we have used the notation $p^x h^y$ for the indices of $\Omega$  
to indicate the relevant $xp-yh$ configuration. 
The additional commutator equations here are not given. 
In order to obtain eigenvalue equations we
need to introduce a cut-off parameter: the GLA~\cite{tom01}, which consists by
applying the Wick's theorem to the 
 $A^{\dagger}_{N+2'}(\beta_{N+2'}(J_1J_2\cdots J_{N+2'})J)$
terms and by neglecting the normal order.
This linearization mechanism generates the additional terms that
 convert the commutator chain in the corresponding eigenvalue equation,
as can be obtained by taking the expectation value of the linearized 
Eqs.~(\ref{eq.9}) and~\ref{eq.9a}) between the vacuum and the model states. 

Using the anticommutation relations and the Wick's algebra,
one verifies easily that $H$ can only connect states that differ by 
$1p1h$.
The eigenvalue equation, Eq.~(\ref{equ.2.6b}),   
at the second-order linearization level 
is given by Eq.~(\ref{equ.3.0})      
where the subscripts referring to 
particle-hole configurations were not written explicitly but are understood.  
Note that in Eq.~(\ref{equ.3.0}) $\Omega_{\alpha p^N\ \beta'' p^{N+2}h^2}$
= $\Omega_{\alpha''p^{N+2}h2\ \beta p^N} = 0$.

\begin{equation}\label{equ.3.0}    
\begin{array}{l}
\displaystyle{ \sum_{\beta\ \beta'\ \beta'' }}   
\left ( \begin{array}{ccc}
 \Omega_{\alpha \beta }-E\delta_{\alpha\beta}  &  \Omega_{\alpha\beta'} & 0 \\
 \Omega_{\alpha' \beta } & \Omega_{\alpha' \beta'}-\delta_{\alpha'\beta'} & \Omega_{\alpha' \beta''} \\ 
    0 & \Omega_{\alpha'' \beta'} &  \Omega_{\alpha'' \beta''}-E\delta_{\alpha''\beta''} 
\end{array} \right)  \\ 
\end{array} 
\displaystyle{ =} 
\begin{array}{c} 
 \\
0   \\
 \\ 
\end{array}
\end{equation}

The self-consistent method of solving Eq.~(\ref{equ.3.0}) is given in 
detail in Ref.~\cite{tom02}. Here, we mention among others that   
in solving Eq.~(\ref{equ.3.0}) the 
two-body interactions of $H$ automatically generates nonlocal
three-, four-interactions and so on.  

The diagonalization of Eq.~(\ref{equ.3.0}) can be performed only if
one can calculate the many-body matrix elements.
Calculations are feasible with the use of the Wick's algebra.
However the number of terms to be evaluated increase exponentially and
calculations are very slow.    
In this work, we perform calculations by using the CFT   
of Ref.~\cite{tom01,tom02,tom07}.    
We believe that with the mastering of the essence of the CFT, matrix elements 
involving even more complex forms of operators can be easily deduced 
from the results obtained here.  

\section{Transition energies in Oxygen ions}
The eDCM finds applications to the calculation of the transition energies of
the Oxygen ions.
In Table~1 we give the energies for the Hydrogen-like Oxygen.
The energies are calculate solving the Dirac's equation in a central Coulomb potential.
For the $1s_{\frac{1}{2}}$ the calculated energy is compare wit the ionization energy
of Ref.~\cite{ene06}. For the energies of the other levels no experimental energies are
available. 
\begin{table}[htp]
\begin{tabular}{|lcc|}\hline
Orbital  & Energy (eV) & Ref.~\cite{ene06}\\ \hline
1s & -871.5080366004061 & 871.41   \\
2s & -217.9238060431288 &  \\ 
2p-& -217.9234006166900 &  \\
2p & -217.7378068079319 &  \\
3s & -96.83498296247103 & \\
3p-& -96.83420091479068 &  \\
3p & -96.77995294892040 &  \\
3d-& -96.77911072701463 &  \\ 
3d & -96.76111291825633 &  \\
4s & -54.46202520546236 &  \\
4p-& -54.46072932022532 & \\
4p & -54.43917924211600 &  \\
4d-& -54.43737498444262 &  \\
4d & -54.43087287667591 & \\
4f-& -54.42959174919736 & \\
4f & -54.42608486286169 & \\ 
5s & -34.85344986252919 & \\ \hline
\end{tabular}
\caption{Energies of the first 17 levels of the Hydrogen-like
Oxygen. The minus sign designates the $j=l-\frac{1}{2}$ states.}
\end{table}
The energies of the Helium-like Oxygen states are then obtained by solving Eq.~(\ref{equ.3.0}).
The indices $(\alpha,\beta)$ are associated to a two electron
states coupled to a good {\it J} quantum number.
The energies of the first three $J=0^+$ states, obtained by diagonalizing 
a matrix with 55 components, are given in Table~2.
In the Table we give only three components of the 55 eCMWFs associated
to the calculated spectroscopic factors.
\begin{table}[htp]
\begin{tabular}{|lcc|}\hline 
Spectroscopic factor &  Orbital  & Energy (eV) \\\hline
 80\%        &  (1s$_{\frac{1}{2}}$1s$_{\frac{1}{2}})^0$           &
-1722.6 \\
 87\%       &  (1s$_{\frac{1}{2}}$2s$_{\frac{1}{2}})^0$           &
-1133.1  \\
 79\%        &  (1s$_{\frac{1}{2}}$3s$_{\frac{1}{2}})^0$           &
-1047.5 \\\hline
\end{tabular}
\caption{The first three levels of Helium-like $O^{6^+}$ with J=$0^+$ and the associated 
spectroscopic factors.}
\end{table}

The energies of Lithium-like states are then obtained by solving Eq.~(\ref{equ.3.0}).
The indices $(\alpha,\beta)$ are associated to a three electron
states coupled to a good {\it J} quantum number.
The energies of the first three $J= \frac{3}{2}^-$ states, obtained by diagonalizing 
a matrix with 350 components, are given in Table~3 together with the
associated spectroscopic factors.

\begin{table}[htp]
\begin{tabular}{|lcc|}\hline
Spectroscopic factor &  Orbital  & Energy (eV) \\ \hline
 92 \%        & $(1s_{\frac{1}{2}}(1s_{\frac{1}{2}}2p_{\frac{1}{2}})^1)^{\frac{3}{2}}$  &
-1949.2 \\
 75 \%        & $(1s_{\frac{1}{2}}(1s_{\frac{1}{2}}3p_{\frac{1}{2}})^1)^{\frac{3}{2}}$  &
-1849.1   \\
99 \%        &  $(1s_{\frac{1}{2}}(1s_{\frac{1}{2}}3p_{\frac{3}{2}})^1)^{\frac{3}{2}}$  & 
-1839.7
\\ \hline
\end{tabular}
\caption{The first three levels of Lithium-like $O^{5^+}$ J=$\frac{3}{2}^-$ and the associated 
spectroscopic factors.}
\end{table}

In order to calculate the transition energies of the Beryllium-like Oxygen
we assume the first 1s$_\frac{1}{2}$ shell full and we diagonalize
Eq.~(\ref{equ.3.0}) with the indices $(\alpha,\beta)$ running
over the unoccupied single particle states and the indices $(\alpha',\beta')$  
over the 1s$_\frac{1}{2}$ closed shell.
The resulting energies for the three $J=1^-$ states obtained by diagonalizing
a matrix of order 750, are given in Table~4 together with the relative 
spectroscopic factors.
\begin{table}[htp]
\begin{tabular}{|lcc|}\hline
Spectroscopic factor &  Orbital  & Energy (eV) \\ \hline
 99\%        &  (2s$_{\frac{1}{2}}$2p$_{\frac{1}{2}})^1$         &
-516.1 \\
 98\%        &  ((2s$_{\frac{1}{2}}$2s$_{\frac{1}{2}})^1$
(2p$_{\frac{3}{2}}$1s$^{-1}_{\frac{1}{2}})^1)^1$                       &
-434.6 \\
 63\%        &  (2s$_{\frac{1}{2}}$2p$_{\frac{1}{2}})^1$         &
-434.3 \\ \hline
\end{tabular}
\caption{The first three levels of Beryllium-like $O^{4^+}$ J=$1^-$ and the associated 
spectroscopic factors.}

\end{table}

In order to calculate the transition energies of the Boron-like Oxygen
we assume the 1s$_\frac{1}{2}$ shell full and we diagonalize
Eq.~(\ref{equ.3.0}) with the indices $(\alpha,\beta)$ running
over the unoccupied single particle states and the indices $(\alpha',\beta')$  
over the closed shell.
The resulting energies for the three $J=0^+$ states obtained by diagonalizing
a matrix of order 614, are given in Table~5 together with the relative 
spectroscopic factors.
\begin{table}[htp]
\begin{tabular}{|lcc|}\hline
Spectroscopic factor &  Orbital  & Energy (eV) \\\hline
 15 \%        &  $(2s_{\frac{1}{2}}(1p_{\frac{1}{2}}4s_{\frac{1}{2}})^1)^{\frac{1}{2}}$    & -5428.1 \\
 37 \%        &  $(2p_{\frac{1}{2}}(3p_{\frac{3}{2}}2p_{\frac{3}{2}})^1
(3p_{\frac{1}{2}}1s^{-1}_{\frac{1}{2}})^1)^{\frac{1}{2}}         $            & -3897.3 \\
 54 \%        &  $(2p_{\frac{1}{2}}(2p_{\frac{1}{2}}3p_{\frac{1}{2}})^1
(2p_{\frac{1}{2}}2s^{-1}_{\frac{1}{2}})^0)^{\frac{1}{2}}     $            & -3734.0 \\\hline
\end{tabular}
\caption{The first three levels of Boron-like $O^{3^+}$ J=$\frac{1}{2}^+$ and the associated 
spectroscopic factors.}
\end{table}

\section{Excitation of the positron continuum}
\subsection{Non-linear realization of the IS theory}  

The knowledge of the theoretical and experimental
mass-dependence (MS) of selected atomic transitions and the theoretical calculations 
of the volume effects (FS) gives the possibility to have a determination 
of the mean-square nuclear radii of short living isotopes~\cite{ott88}.  
 Recent values for the nuclear charge radii of short-lived 
lithium and helium isotopes have been obtained from measurements 
performed at GSI, Vancouver~\cite{ewa05},
Argonne~\cite{wan04}. The measurements of the $2^2S_{1/2}\to3^2S_{1/2}$,
of the $2^2S_{1/2}\to2^2P_{3/2}$, and $2^2S_{1/2}\to2^2P_{1/2}$ transitions
together with the recently performed calculations~\cite{dra02} of the same transitions
in lithium and helium atoms were in fact used to extract the difference of
the nuclear charge radii of the short-living isotopes from the
charge radius of the stable isotope.
In this paper we propose to reevaluate the MS and the FS in a non perturbative 
approximation based on the application of the eDCM.

 We start to calculate the energies of the lithium atoms
by diagonalizing Eq.~(\ref{equ.3.0}) in a base formed by three
electrons in the (s,p,d) single particle states
which interact with the BOV states formed by exciting the $e^--e^+$
continuum states.
Results of this calculation for the 2s and 3s states are shown in table~I.

\begin{table}[htp]
\begin{tabular}{|lcc|}\hline
  References   &method& $1s^22s$ energies in au \\ \hline
  Chung~\cite{tkc91}& FCPC & -7.47805797(9)\\
  McKenzie and Drake~\cite{mkz93}& HR & -7.478060326(10)\\ 
  Yan and Drake~\cite{dra02} & HR &  -7.47806032310 (31) \\
  Tomaselli & eDCM & -7.478060733\\ \hline
  References  & method & $1s^23s$ energies in au \\ \hline
  Yan and Drake(02)~\cite{dra02}& HR & -7.354098422249(18)\\
  King~\cite{kin91}& HR & -7.354076\\
  Wang et al.~\cite{wan92} & -  & -7.5440980 \\
  Tomaselli & eDCM & -7.35409801\\ \hline
\end{tabular}
\caption{Calculated energies of the $1s^22s$ and the $1s^23s$ in different models.}
\end{table}

According to Ref.~\cite{ott88} in order to evaluate the Mass Shift (MS) we have 
to add to the eigenvalue 
equation the additional term: $\nabla_i.\nabla_j$ and to rescale energies and distance
with the reduced mass of the electron.
The matrix element of the $\nabla_i.\nabla_j$ can be calculated as in Ref.~\cite{tom05}
while the rescaling of the energies can be obtained
 by adding a $\vec{R}_{\mathrm{nucl}}.\vec{r_i}^{\mathrm{electron}}$ term 
to Eqs.~(\ref{eq.9}) and (\ref{eq.9a}) 
 and to re-diagonalize the matrix given in Eq.~(\ref{equ.3.0}).  
The correlations of the nucleus, which influence via the additional
matrix elements given above, are in general approximated by a non relativistic
perturbative calculations~\cite{dra02}.

The FS term~\cite{bri58}
factorize into a constant 
\begin{equation}\label{equ.5.1}
C=\frac{2\pi Z}{3}[ \langle \delta{\vec{r}}\rangle_{3s}-\langle \delta{\vec{r}}\rangle_{2s}]   
\end{equation}
where the term $\langle\delta{\vec{r}}\rangle$ denotes the expectation value of the electron density
at the nucleus multiplied by the isotopic variation of the charge radius. 
The polarizability of the nucleus, which influences the calculation of this constant,
 has been evaluated relative to the polarization of deuterium~\cite{pac96}.
Since the FS is generally calculated in the point nucleus approximation,
calculations performed within the DCM (nucleus) and the eDCM (electrons) correlation models 
could give a better insight in the FS calculation.

Calculation of the IS for the isotopes of Lithium and Helium are under present
calculation and will be reported soon.
\subsection{Transition energies in Lithium-like $^{235}$U.}
The 2s-2p transition of Lithium-like $^{235}$U is calculated in the eDCM.
The result is given in Table~7 and compared with the QED calculation
of Yerokhyn~\cite{yer00} and with the experimental result~\cite{bei00}.
By using the resulting eCMWFS for the $2s_{\frac{1}{2}}$ and $2p_{\frac{1}{2}}$
we can calculate the hyperfine splitting (HF) of the two states.
The calculation are performed by coupling the three electron wave functions
to the ground state wave function of $^{235}$U.
For the nuclear ground state
wave function we use a DCM which reproduce well within a large dimensional space
the nuclear energies and moments of the $2f\frac{7}{2}$ valence neutron.
Detailed calculation will be reported soon.

\begin{table}[tbp]\label{b-I}
\begin{tabular}{|lc|}
\hline
Transition   & 2s-2p  \\ \hline
eDCM         & 280.33  \\ 
Yerokhyn-Shabaev & 280.44(20)\\
Experiment & 280.645(15) \\ \hline
\end{tabular}
\caption{2s-2p Transition in Lithium-like $^{235}$U}
\end{table}

\section{Conclusion and Outlook} 
The transition energies of the Oxygen ions are calculated within a microscopic
cluster model. The model is derived from the unitary
operator model $e^{iS}$ which is used to generate the eCMWFs.  
The amplitudes of the model of the eCMWFs are calculated by using the EoM method.
The modifications caused to the energy transitions by a time dependent laser beam
can be simply evaluated by modifying the EoM. 
For few electron atoms like Helium and Lithium the EoM are extended to include
the BOV excitations (excitation of the positron-continuum). 
The calculated eCMWFs together with the nuclear CMWFs of the different
isotopes of Helium and Lithium allow a non-perturbative evaluation of the 
MS and FS of the IS theory.
The influence of this new evaluation method on the charge radii 
of the Helium and Lithium isotopes is under present investigation.
An open point in the presented calculation is the determination of the error 
of the calculated transition energies.
In performing structure calculations we have used the single spinor energies obtained from
the solution of the Dirac's equation (see Eq.~(\ref{equ.2.0}).
Better energies can be obtained by using the Harthee-Fock method.
The approximation we have used gives to the calculated energies
 an error that can vary depending from the electron energy considered from 0.1 to few percent.  
A better estimation of the errors could however be given, as suggested by Drake, by evaluating
elementary excitation processes in light atoms like Hydrogen.
For this purpose we are investigating the two photon transitions
in Hydrogen. This would allow to establish a connection between the 
present non perturbative method and the QED perturbation theory.

\end{document}